\documentclass[sigconf]{acmart}

\AtBeginDocument{%
  \providecommand\BibTeX{{%
    \normalfont B\kern-0.5em{\scshape i\kern-0.25em b}\kern-0.8em\TeX}}}

\copyrightyear{2022}
\acmYear{2022}
\setcopyright{rightsretained}
\acmConference[WWW '22]{Proceedings of the ACM Web Conference
2022}{April 25--29, 2022}{Virtual Event, Lyon, France}
\acmBooktitle{Proceedings of the ACM Web Conference 2022 (WWW '22),
April 25--29, 2022, Virtual Event, Lyon,
France}\acmDOI{10.1145/3485447.3512271}
\acmISBN{978-1-4503-9096-5/22/04}

\acmConference[WWW'22]{The ACM Web Conference
Proceedings}{April 25--29, 2022}{Virtual Event, Lyon,
France}
\acmPrice{15.00}
\acmISBN{978-1-4503-9096-5/22/04}

\acmSubmissionID{w12fp3048}

\usepackage{subcaption}
\usepackage{algorithm}
\usepackage{algorithmic}

\begin{document}

\title[Construction of Large-Scale Misinformation Labeled Datasets from Social Media Discourse]{Construction of Large-Scale Misinformation Labeled Datasets from Social Media Discourse using Label Refinement}

\author{Karishma Sharma}
\email{krsharma@usc.edu}
\orcid{0000-0001-6825-5876}
\affiliation{%
  \institution{University of Southern California}
  \country{USA}
}

\author{Emilio Ferrara}
\affiliation{%
  \institution{University of Southern California}
  \country{USA}}
\email{emiliofe@usc.edu}

\author{Yan Liu}
\affiliation{%
  \institution{University of Southern California}
  \country{USA}
}
\email{yanliu.cs@usc.edu}

\begin{abstract}
  Malicious accounts spreading misinformation has led to widespread false and misleading narratives in recent times, especially during the COVID-19 pandemic, and social media platforms struggle to eliminate these contents rapidly.
 This is because adapting to new domains requires human intensive fact-checking that is slow and difficult to scale.
 To address this challenge, we propose to leverage news-source credibility labels as weak labels for social media posts and propose model-guided refinement of labels to construct large-scale, diverse misinformation labeled datasets in new domains. The weak labels can be inaccurate at the article or social media post level where the stance of the user does not align with the news source or article credibility. We propose a framework to use a detection model self-trained on the initial weak labels with uncertainty sampling based on entropy in predictions of the model to identify potentially inaccurate labels and correct for them using self-supervision or relabeling. The framework will incorporate social context of the post in terms of the community of its associated user for surfacing inaccurate labels towards building a large-scale dataset with minimum human effort. To provide labeled datasets with distinction of misleading narratives where information might be missing significant context or has inaccurate ancillary details, the proposed framework will use the few labeled samples as class prototypes to separate high confidence samples into false, unproven, mixture, mostly false, mostly true, true, and debunk information. The approach is demonstrated for providing a large-scale misinformation dataset on COVID-19 vaccines. \textbf{Dataset:} \url{https://github.com/USC-Melady/Constructing-Misinformation-Datasets-WWW-2022}.
\end{abstract}



\maketitle

\section{Introduction}
In recent times, malicious and coordinated promotion of misinformation coupled with uncertainties in real-world events, have sparked a plethora of \emph{false} and \emph{misleading} information on social media platforms \cite{sharma2020identifying}. Social media platforms struggle to eliminate these contents effectively and in a timely manner, and are recently attempting to solve the problem through more crowdsourced approaches to misinformation identification \cite{nprbirdwatch}. Twitter introduced \emph{`Birdwatch'} in 2021, which allows people to identify tweets they believe are misleading and provide notes with additional context. This is an effort to respond more quickly to diverse false and misleading claims present on the platform. However, this comes with the biggest challenge for Twitter in ensuring that Birdwatch itself does not become prey to malicious coordinated operations. Secondly, due to ideological biases about the truth it is a challenge to build reliable consensus from platforms like Birdwatch \cite{nprbirdwatch}. 

The central challenge in timely misinformation detection, mitigation, and analysis is the difficulty in obtaining labeled misinformation datasets at scale, especially in new domains. Moreover, diverse and evolving false and misleading information based on changing real-world events, constantly surface on social media \cite{sharma2019combating}. In the literature, the two primary approaches to constructing misinformation datasets are based on either collecting available \textbf{fact-checked claims} from organizations like Snopes, PolitiFact, etc. \cite{ma2016detecting, shu2020fakenewsnet}, or utilizing \textbf{news-source credibility} labels based on reliable or unreliable sources listed by fact-checking organizations \cite{zhou2020recovery,sharma2022election}. The former approach suffers from claim selection bias with a slow and not scalable, human intensive fact-checking process, unsuitable for timely identification in new and evolving domains. The latter approach allows for more diverse, large-scale misinformation labeled social media posts, from a handful of analyzed news sources but can contain inaccurate labels in the dataset.


\textbf{Proposed Approach:} We address the above shortcomings with an alternate approach to construct misinformation datasets. We propose to utilize news-source credibility labels as weak labels for social media posts, and use \emph{model-guided} refinement of labels to construct large-scale, diverse misinformation datasets in new domains. The news-source credibility based labels can be inaccurate at the article or social media post level when the stance of the user does not align with the news-source or article credibility. Therefore, for label refinement, we propose to use self-supervision from \emph{any} generic misinformation detection model, with social context modeling of the social media posts. In this framework, we use a misinformation detection model trained on the initial weak labels, with uncertainty sampling based on entropy in predictions of the model to identify potentially inaccurate labels and correct for them using self-supervision or relabeling. In addition, we incorporate the social context of the post in terms of the community of its associated user to model user credibility and stance in the discourse. The model-guided refinement is used to surface inaccurate labels iteratively, and minimize human labeling efforts, enabling timely scaling to large misinformation datasets with greater coverage.

The model-guided confidence in the labels is used to filter out or correct inaccurate weak labels, and the resulting dataset of social media posts with its engagements are labeled as misinformation/reliable, and associated with a model confidence in its label. The misinformation can be further segregated at finer-grained labels (such as false, mixture, true \cite{sharma2019combating}) which can be obtained after label refinement with a \emph{semi-supervised classification} setup \cite{kipf2016semi,xie2020unsupervised} in the proposed approach. Specifically, to provide finer-grained labels, we use the few human labeled examples as class prototypes to separate high confidence examples into \emph{false, unproven, mixture, mostly false, mostly true, true, and debunk} information. More details on the labels and annotation guidelines are discussed later. The approach is demonstrated and applied for constructing and providing the research community with a large-scale public misinformation dataset on COVID-19 vaccines. 


\textbf{Contributions:} Our contributions developed in this work are:
\begin{itemize}
    \item Model-guided label refinement approach for timely construction of large-scale misinformation datasets.
    \item Label annotation guidelines and flexible framework that can generalize to other misinformation domains.
    \item Evaluation and construction of public misinformation dataset on COVID-19 vaccine social media data from Twitter.
\end{itemize}

In the following sections, we discuss the challenges in misinformation dataset construction, limitations of existing methods, related works, and the proposed approach and experiments, and conclude with a discussion of limitations and future work.

\section{Related Work}

\textbf{Misinformation datasets.} Misinformation referring to false and distorted facts on social media has been addressed in numerous studies. A review of misinformation detection, mitigation techniques, and related datasets and tasks is comprehensively surveyed in \cite{sharma2019combating}. The construction of misinformation datasets is a central task to enable research on misinformation detection, mitigation and analysis. Existing misinformation datasets are either general, such as over a specific time period \cite{ma2016detecting} and cover content, social media engagements, and temporal features \cite{shu2020fakenewsnet}, or topic-specific datasets such as on the Syrian war \cite{salem2019fa}. The label scheme of datasets and the type of information collected vary based on the specifics of the task. For instance, for claim verification with external knowledge, datasets include content and evidence collected from the web that support or refute claims in the content \cite{popat_where_2017}. The general detection task requires learning discriminative classifiers for misinformation claims, and usually includes content and its social media engagements, and the labels depend on the distinction made during data collection e.g. fake/real news \cite{dai2020ginger} unreliable/reliable \cite{zhou2020recovery} rumors/non-rumors \cite{ma2016detecting}. A comprehensive summary of several popular datasets in terms of their label classes and features is available in \cite{sharma2019combating,shu2020fakenewsnet}.

\textbf{Misinformation detection.} Misinformation detection relies on learning discriminative features from labeled datasets, often utilizing the propagation features, content features, and account features \cite{qian2018neural,ma2016detecting}. \citeauthor{wang2020weak} \cite{wang2020weak} in addition use weak supervision from user's reports to augment labeled misinformation datasets with unlabeled examples for misinformation detection. \citeauthor{shu2020detecting} \cite{shu2020detecting} use weak social supervision to similarly improve misinformation detection, i.e., where social media engagements are abundant but labeled misinformation content is not, modeling the interactions between social media users and contents to improve discrimination of misinformation content. Both these works are similar in flavor, in terms of \emph{augmenting} the misinformation labeled datasets with auxiliary information to improve detection. In our work, we address how to scale the construction of misinformation labeled datasets using news-source credibility as initial weak labels.

\textbf{Label refinement.} Label noise in real-world data is common, and there are many different approaches to detect, remove, or correct it, which are relevant to this work \cite{sharma2020noiserank,kremer2018robust,arazo2019unsupervised}. Some works use local label inconsistencies in the feature space for detection \cite{sharma2020noiserank}, others utilize the training loss of deep neural network classifiers on the dataset to filter examples with high training loss in early epochs as noisy \cite{arazo2019unsupervised}, or utilize entropy or variance in classifier predictions \cite{frenay2013classification}. Other works focus on making classifiers more robust to label noise in datasets \cite{song2020learning}. Active learning works address selection of instances from unlabeled or labeled datasets that are most useful to get human labels for to learn better models from the data, but depend on the presence of an `oracle' i.e., human labeler, and utilize the expected model change from human labeling to select which instance to pick \cite{kremer2018robust}. 
Here, we propose to additionally incorporate social context in label refinement, since in social media applications, the structures and context of social media users, as we show, provides relevant, complementary signals.



\section{Data Collection}
\label{sec:data}
We collected social media posts on COVID-19 vaccines using Twitter's streaming API from December 9, 2020 - Feb 24, 2021 with keywords related to the vaccines ("Vaccine", "Pfizer", "BioNTech", "Moderna", "Janssen", "AstraZeneca", "Sinopharm"). The stream fetches a $\sim1\%$ sample of all tweets containing at least one of the keywords from the platform in real time. The data collection period started just prior to the first Emergency use authorization of Pfizer-BioNTech COVID-19 vaccine in the U.S. 
The dataset contains \textbf{4,764,701} unique user accounts with \textbf{15,158,523} collected tweets.





\subsection{News-source credibility labels} 

Previous works that use news-source credibility for misinformation labeling, include news sources analyzed by different fact-checking organizations \cite{Bozarth_Saraf_Budak_2020}. \citeauthor{Bozarth_Saraf_Budak_2020} \citeyear{Bozarth_Saraf_Budak_2020} found that differences in lists based on the fact-checker it  is compiled  from affect  prevalence,  but  not  the temporal  trends  or  differences  in  narratives of  misinformation vs.  legitimate contents labeled by these methods. In this work, more than prevalence, we are interesting in curating news sources to provide weak labels covering a diverse set of 
possible misinformation found in social media posts. 

Therefore, we compile news-source credibility labels from multiple fact-checking resources, to encompass a wide range of low-credibility news sources. Following \cite{sharma2022election}, we collect lists of unreliable and conspiracy news sources from three fact-checking resources: Media Bias/Fact\footnote{\href{https://mediabiasfactcheck.com/}{https://mediabiasfactcheck.com/}}, NewsGuard\footnote{\href{https://www.newsguardtech.com/covid-19-resources/}{https://www.newsguardtech.com/covid-19-resources/}},
and  Zimdars \cite{zimdars2016false}\footnote{\href{https://21stcenturywire.com/wpcontent/
uploads/2017/02/2017-DR-ZIMDARS-False-Misleading-
Clickbait-y-and-Satirical-\%E2\%80\%9CNews\%E2\%80\%9DSources-
Google-Docs.pdf}{https://21stcenturywire.com/wpcontent/
uploads/2017/02/2017-DR-ZIMDARS-False-Misleading-
Clickbait-y-and-Satirical-\%E2\%80\%9CNews\%E2\%80\%9DSources-
Google-Docs.pdf}}. NewsGuard maintains a repository of news publishing sources that have actively published false information during the COVID-19 pandemic. The listed sources from NewsGuard, accessed on September 22, 2020 are included, along with low and very-low factual reporting listed as questionable from Media Bias/Fact Check, and sources tagged with unreliable or related labels and conspiracy/pseudoscience from Zimdar's list. List of reliable sources\footnote{\href{https://en.wikipedia.org/wiki/Wikipedia:Reliable_sources/Perennial_sources}{\url{https://en.wikipedia.org/wiki/Wikipedia:Reliable_sources/Perennial_sources}}} \cite{sharma2022election}, covering high factual sources is also collected for obtaining the weak labels. In total, we obtained \textbf{1380} unreliable (or conspiracy) and \textbf{124} reliable sources. This choice of lists provides informative weak labels (ref. Section \ref{sec:challenges} and \ref{sec:experiments}) but can be replaced or updated with other resources on news-source credibility analysis as needed.

\subsection{News-related tweet cascades} 

On social media, content propagates through the network when accounts engage with posts by re-sharing (retweet), replying (reply tweets), quoting (quote tweets are retweets without a comment). A reply tweet can also be retweeted or quoted, and likewise for quote tweets. Therefore, source posts receive direct and subsequent indirect engagements through propagation over the network. This flow of information is referred to as an `information cascade' \cite{yang2010modeling} or tweet cascade. We represent it as a sequence of tweets, ordered by their time-stamp. The source post is the first tweet in the cascade. Formally, a cascade can be represented as follows with the user (u), tweet (tw), and temporal (t) features of when the users posted the engagements \cite{ruchansky2017csi}, 
\begin{equation}
    C_j = [(u_1, tw_1, t_1), (u_2, tw_2, t_2), \cdots (u_n,tw_n, t_n)] 
\end{equation}

\textbf{Extracting tweet cascades.} To extract the content cascades from the collected data, we use the retweet/reply/quote links between the tweets available from its metadata, and construct a directed graph of the tweets. We find the weakly-connected components of this graph, and each corresponds to one tweet cascade \cite{sharma2020covid}. \textbf{Weak labels using news-sources.} The cascade is weakly labeled based on news-source credibility lists if the source post references one of the news sources. The news-source label (unreliable, conspiracy, reliable) is assigned to the cascade as its weak label.

We extracted tweet cascades from the collected Twitter data stream sample, keeping 490,638 user accounts that have at least 5 collected tweets in the sampled stream. 
The total tweets for these accounts is 9M. We weakly labeled the tweets as mentioned, and obtained \textbf{10,377} reliable cascades, and \textbf{4,267} unreliable or conspiracy cascades. These 14.6k cascades with weak labels will be used to construct the misinformation dataset as described in later sections. 


\begin{figure}[t]
    \includegraphics[width=1.1\columnwidth]{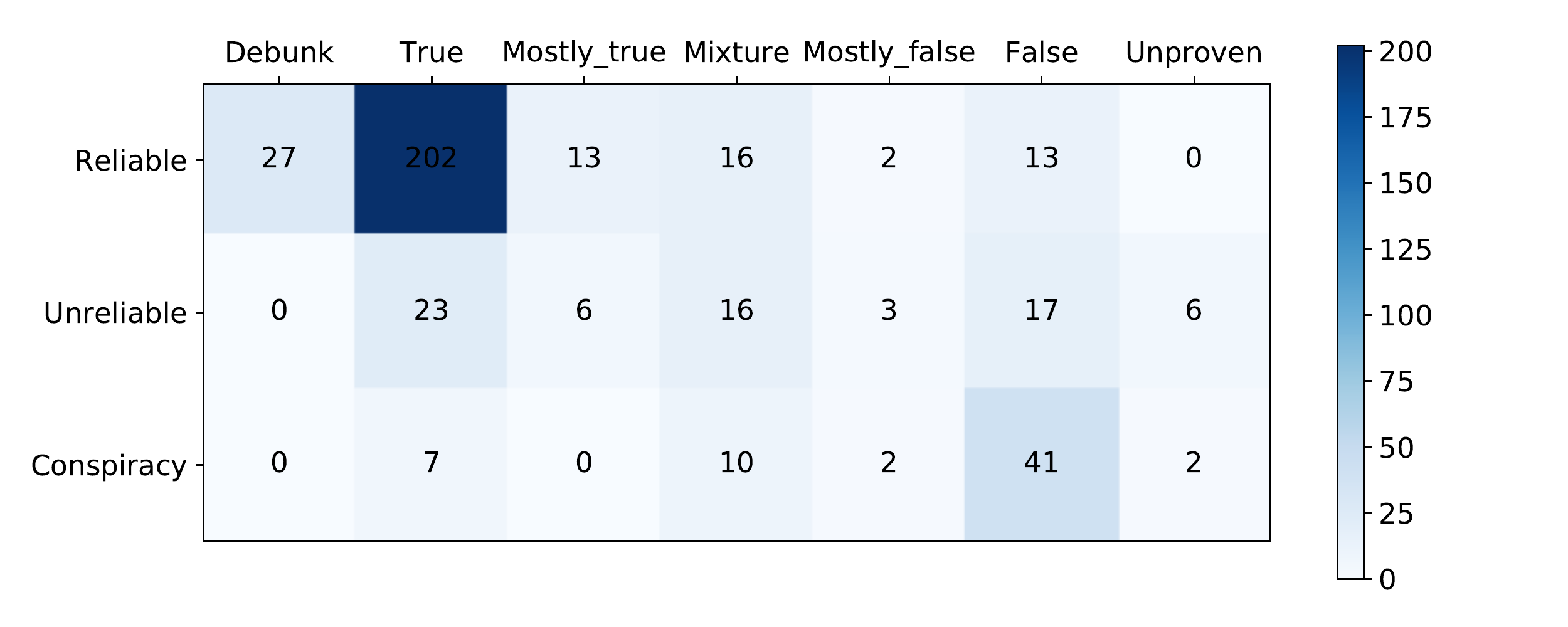}
    \caption{News-source credibility labels correlation with fact-checked claim based labels on validation and test tweets.}
    \label{fig:correlation_ns_label_human_label}
\end{figure}

\section{Challenges in Misinformation Labeling}
\label{sec:challenges}

\subsection{Existing approaches and limitations}

Existing works apply two primary approaches to construct misinformation labeled datasets, either from claims verified by fact-checking organizations, or using credibility of the sources publishing the content. Both approaches suffer from drawbacks. We describe the approaches and summarize the drawbacks below. \textbf{Fact-checking based labeling.} One approach to collecting labeled misinformation contents, in the form of news articles, claims, or social media posts, is from contents verified by fact-checking websites (e.g., Snopes, PolitiFact). This approach is frequently used to construct datasets with few hundred or thousand labeled misinformation contents  \cite{ma2016detecting,shu2020fakenewsnet,kwondata}. Then, for the fact-checked claims, social media engagements related to it are collected by search for content keywords using social media API's (e.g. Twitter search API). The matched social media posts containing these keywords are inspected to determine if they are relevant to the content  \cite{kwondata}, or the search keywords are refined until reasonably relevant matches of social media engagements are collected \cite{ma2016detecting}. \textbf{News credibility-based labeling.} The other approach used for misinformation labeling is based on credibility of news sources \cite{Bozarth_Saraf_Budak_2020}. Social media posts referencing content from any of these sources is labeled based on the source credibility to provide a dataset of unreliable and reliable contents. This is used frequently to identify misinformation posts from social media discourse for timely analysis in new domains \cite{zhou2020recovery,singh2020first}. The drawbacks are summarized as follows.

\begin{itemize} 
    \item \emph{Fact-checking based labeling.} It can have claim \emph{selection bias}, since fact-checkers usually select claims to verify based on relevance or popularity (e.g., PolitiFact\footnote{\url{https://www.politifact.com/article/2018/feb/12/principles-truth-o-meter-politifacts-methodology-i/\#How\%20we\%20choose\%20claims}}), which can limit the diversity of collected claims for detection, and bias the analysis. Plus it is slow, human-intensive, and less scalable. 
    \item \emph{News source credibility-based labeling.} It scales to many social media posts using a handful of unreliable (questionable) and reliable sources, resulting in more diversity in claims, but has inherent label noise at the article or social media post level. Therefore, it can only provide weak labels.
\end{itemize}

\subsection{Correlation between news-source credibility and fact-checked claim based labels}
\label{sec:correlation_sets}
First, we analyze the correlation between the labels from the two approaches. We collected fact-checked claims from Snopes.com\footnote{\url{https://www.snopes.com/tag/covid-19-vaccine/}} and NewsGuard\footnote{\url{https://www.newsguardtech.com/special-reports/special-report-top-covid-19-vaccine-myths/}} on COVID-19 vaccines. For each fact-checked claim, we find tweet cascades that discuss the claim by searching for text matches to words related to the claim. E.g. ``Myth: The COVID-19 vaccine will use microchip surveillance technology created by Bill Gates-funded research." We search for source tweets with words ``chip", ``microchip", ``surveillance" for matches. If nothing is found, we refine the search with``gates" and sample to check for matches. 

NewsGuard provides only Myths (false claims), while Snopes provides varying factuality labels (true, mostly true, mixture, false etc.). From the Snopes collection tagged as COVID-19 vaccines, we obtained the claims labeled as one of these types (tagged as fact-checked) or labeled as news articles (AP news, The Conversation). Associated Press or AP news are rated as very high factual reporting and least biased by Media Bias/Fact Check. Therefore, we take claims from AP News as reliable. For more reliable claims, we also directly crawl the websites of AP news and NPR news (same factual and bias rating as AP news) for news articles (extracting the article heading, claim/short description, date) using python web-scraping. Snopes, AP news and NPR news together give us 400 claims, which we sample to find matching tweet cascades. We inspected the source tweet and labeled it based on the stance of the tweet to the fact-checked claim or reliable news article as `true, mostly true, mixture, false, mostly false, unproven, debunk', similar to Snopes.

We found 256 tweet cascades to label based on the Snopes fact-checks and News articles. This forms our \textbf{evaluation test set of tweets with human expert ground-truth labels}. To additionally construct a \textbf{validation set} of human labeled tweets, we used stratified sampling of 150 additional tweets from the 14k tweet cascades and labeled them based on similar annotation guidelines as Snopes, described in the next subsection.

In Fig~\ref{fig:correlation_ns_label_human_label}, we compare the news-source credibility based labels (unreliable/conspiracy/reliable) with the inspected fact-checked claim based labels for the human labeled tweets in the evaluation test and validation sets mentioned above. Overall, the news credibility labels appear to be well correlated with actual human labels. Individual inaccuracies can still exist, but with the large-scale weakly-labeled data smoothing out individual errors, we could learn to refine the weak labels to construct the misinformation dataset. 

\subsection{Annotation scheme and guidelines}
\label{sec:anno_main}
Labeling misinformation is already challenging, more so because misinformation is not easy to specify \cite{sharma2019combating}. It can lie on a spectrum of truth, including false, conspiracy \cite{ferrara2020characterizing}, and misleading or \emph{distorted} information such as missing or misleading contexts or mixture \cite{wardle2017fake,lewandowsky2021covid}. We find that fact-checking organization Snopes uses a well-defined label schema that is \textbf{general enough to fit any domain}, and yet manages to \textbf{cover all types} and nuances of distortion we found upon examining tweets in the vaccines dataset, and generally in the literature \cite{sharma2019combating}. Snopes includes several labels to cover the varying degree of truth and other deceptive tactics like miscaptioned, misattributed, scam. We work with the 6 most relevant Snopes categories, and add the `Debunk' category based on what we observed in tweets. These cover even very specific types of anti-vaccine misinformation and science distortions \cite{lewandowsky2021covid}.

The label scheme is proposed below, derived from Snopes and tweet inspection. We refine the label definitions to make the distinction between them and its coverage explicitly clear based on the inspected tweet data, for labeling social media posts based on their factualness. \textbf{Guideline:} Label the tweet based on what the tweet is trying to say or claim, and how factual its claim is.
Choose one of the below labels for the tweet:

\begin{itemize}
    \item True: Primary elements of the claim are demonstrably true.
    \item Debunk: Tweet calls out or debunks inaccurate information.
    \item Mostly true: Primary elements of a claim are demonstrably true, but some of the \emph{ancillary details} surrounding the claim \emph{may be inaccurate}.
    \item Mixture: Claim has \emph{significant elements of both} truth and falsehood (including for e.g. \emph{significant missing context} or \emph{misleading} which might cause one to be misled about truth).
    \item Mostly false: Primary elements of a claim are false, but ancillary details may be accurate.
    \item False: Primary elements of a claim are false or conspiratorial.
    \item Unproven: Insufficient evidence that it is true, but for which declaring it false would require a difficult (if not impossible) task of proving a negative.
\end{itemize}

  

We evaluated the label scheme and guidelines on a random 200 sample subset from the collected tweet cascades. We compute the inter-annotator credibility for the tweets between two annotators, one graduate non-native English speaker familiar with misinformation research vs. one undergraduate native English speaker not familiar with the research. The agreement is moderate if we consider across the 7 label categories (0.61 Cohen's kappa), and substantial (0.77 Cohen's kappa) if binarized as (true, debunk, mostly true) vs. (mixture, mostly false, false, unproven) as high-level abstractions. Both annotators followed the same guideline and instructions with typical and difficult examples (noted in Appendix \ref{sec:appx_anno}).\footnote{This label scheme and instruction set was finalized after two iterations. At first, we started with 11 labels with vaccine  specific instructions e.g. potential to mislead (can be misinterpreted as unsafe), missing context, etc) but found difficulties in making clear distinctions and following the specified instructions. This label scheme was clear to follow and we found that sufficient examples of tricky and typical cases in each category was very helpful to the annotators, who were asked to review the instructions and examples before annotating, and had access to reference it when in doubt.
} 

\begin{figure*}
    \centering
    \includegraphics[width=0.8\textwidth]{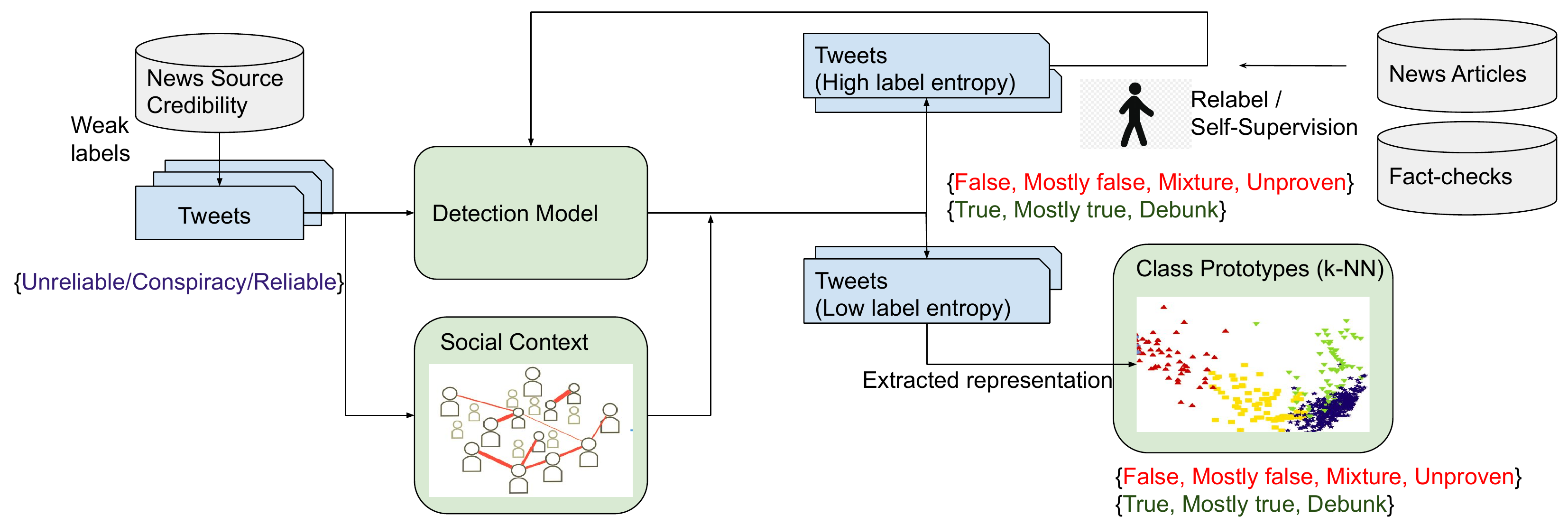}
    \caption{Proposed approach: Model-guided label refinement with self-supervision from a generic misinformation detection model and social context modeling to construct large-scale misinformation labeled social media datasets in a timely manner.}
    \label{fig:pipeline}
\end{figure*}

\section{Model-Guided Label Refinement}

We propose an alternate approach to constructing misinformation datasets at scale, addressing the shortcomings of existing approaches. We propose to use news-source credibility as weak labels and leverage the large-scale weak labeled data with label refinement to construct misinformation datasets minimizing time-consuming, and not scalable human labeling efforts.

In the previous section, we observed that the news credibility labels are correlated overall with actual fact-checked claim based labels, and with the large scale of the weak labeled dataset smoothing out individual errors, we could learn to remove inaccuracies. The inaccuracies in these weak labels arise at two levels: 1) \textbf{Article level.} First, not all contents published by misinformation news sources might contain misinformation, although they tend to be unreliable or repeatedly violate journalistic reporting principles, enough for the source to be included as a low factual reporting source by experts. 
2) \textbf{Tweet level.} Secondly, the weak label can be incorrect based on the stance of the social media post to the content from the news source. The post or tweet might reference content from the source with its supporting viewpoint or restating it as is. Or in other cases,  oppose or distort the content from the source, which would result in a mislabeling at the level of a tweet.


 
In \textbf{Fig.~\ref{fig:pipeline}}, we provide the proposed framework for misinformation dataset construction and labeling in new domains at scale. The weak labels \textbf{unreliable, conspiracy, reliable} on tweet cascades are from news-source credibility. It is utilized to construct the initial dataset with $y \in \{0, 1\}$ as weak labels with 1 as unreliable/conspiracy and 0 as reliable. Our goal is to remove or correct inaccurate weakly labeled instances in the dataset, and output high-level distinctions of misinformation label as 1 and reliable information label as 0, with the model-guided predictions of confidence in the labels.

\subsection{Self-supervision from misinformation detection modeling}
In the proposed framework, we make use of \emph{any} generic misinformation detection model to guide the weak label refinement. Classifiers are often utilized to estimate uncertainty in the instance labels from the loss or model predictions in label noise methods \cite{fefilatyev2012label,arazo2019unsupervised}. In this work, we use entropy of the misinformation detection model predictions to measure closeness from the decision boundary \cite{fefilatyev2012label}. High entropy indicates greater model uncertainty about the label. The entropy $S_i$ in the model predictions for an instance $i$ is defined as follows, where $p(x_i)$ is the vector of predicted probabilities from the detection model, and $k$ is the classes, here for $y \in \{0,1\}$.
\begin{equation}
    S_i = -\sum_{k=1}^{L} p_k (x_i) \log(p_k (x_i))
\end{equation}

 We train the detection model on all weak labeled data, and then filter out high entropy instances. We also filter out tweets with low entropy model predictions if the initial weak label and predicted model label for the tweet are inconsistent with each other. This is for instances where the model is confident in its prediction, either has an incorrect weak label or predicted label. 
 
 With the filtered dataset, we retrain the detection model, and repeat until the model has marginal improvements on a held-out small human-labeled or fact-checked labeled validation set is marginal. This iterative self-training improves the detection model and its signals of the inaccuracies in weak labels. In each iteration, the retrained detection model is applied to all instances in the initial dataset to calculate the entropy scores, and filter for the next iteration as it can now make more informed filtering decisions than those in the previous iterations.
 
\subsection{Social context modeling} In misinformation applications, social media engagements are known to provide useful signals for misinformation detection \cite{shu2020fakenewsnet}. Here, we propose that the social context can also be useful in guiding the construction of misinformation labeled datasets. We describe how \emph{social context can be modeled and leveraged} in this application. 

We incorporate social context of the post using the community of its associated user to model a user account's credibility and stance in the discourse. Social media discourse tends to be segregated into \emph{echo-chambers} of user accounts sharing similar opinions \cite{garimella2018quantifying}. User accounts follow each other based on their interests, and become more exposed to contents that align with their interest and ideologies \cite{sharma2019combating}. The retweet graph between user accounts that retweeted each other's tweets can be used to identify \textbf{user communities}. Retweets are seen as a form of endorsement of content and edges with at least two retweets are retained to capture links of similar interests in the user accounts \cite{garimella2018quantifying}. We identify user account communities from the retweet graph using \emph{Louvain} method \cite{blondel2008fast}.

To leverage the social context, we identify communities that dominantly post or share misinformation sources vs. reliable sources. Several works have found that \emph{informed} and \emph{misinformed}
user accounts exhibit echo-chambers in their network structure \cite{memon2020characterizing,sharma2021covid19vaccine}. For the identified communities, if tweets of user accounts in the community dominantly contain references to misinformation news sources, the communities are likely less to be credible, or are more misinformed. Misinformation communities would involve either malicious groups promoting misinformation, or groups with beliefs that support or are vulnerable to believing and sharing misinformation on the topic of the discourse \cite{memon2020characterizing}. We can thereby leverage this to encode a \textbf{user account's credibility and stance} with respect to misinformation on the topic of the discourse. Here, we denote social context of a tweet as the community of the user account that posted the tweet. Given the community structure, the tweet cascade is detected as possibly mislabeled if,
\begin{itemize}
    \item the user account belongs to an identified dominant misinformation (unreliable/conspiracy) news-source sharing community, but the tweet is weakly-labeled as reliable.
    \item or, if the account is in a dominantly reliable information sharing community, but the weak label is unreliable/conspiracy.
\end{itemize}
For mixed communities with unclear dominant reliable or misinformation sharing patterns, we have no definitive social context for label refinement, and use only the detection entropy. 

\subsection{Iterative label refinement} We jointly use the social context signal and the detection model entropy to guide the identification of post-level mistakes in the weakly-labeled data. The proposed framework (Fig~\ref{fig:pipeline}) is iterative and flexible. We can replace the misinformation detection model with any modeling choice, and use either \textbf{self-supervision and/or human/model based relabeling}. The detection model is first trained by itself with self-supervision from the model predictions. Then the improved detection model signals are jointly combined with social context for further label refinement. The process is iteratively repeated with evaluation of detection model on small held-out human labeled or fact-checked validation set as a proxy for label quality in the large-scale dataset.

The procedure for label refinement from detection model and social context is described in  \textbf{Algorithm~\ref{alg:labelrefinement}}. The subroutine assumes as input the instance (tweet cascade denoted as $x_i$, with its weak label $y_i$), the detection model $M$ trained in the previous iteration, and the social context $S$. Given the model state, we generate three possible \textbf{actions:} (1) RETAIN weak label (2) FLIP weak label (3) QUERY label. Action retain keeps the instance with its weak label in the dataset, flip is model-guided relabeling (without human resource), and query is for active human relabeling of the model suggested instance. If the human resource is not available, then QUERY can be replaced by REMOVAL (discarding the instance due to low confidence in its label or due to contradictory confident signals from detection model $M$ and social context $S$).

The \textbf{states} from the detection model $M$ and social context $S$ are defined as follows, for instance $x_i$,
\begin{itemize}
    \item M-lc: If high-entropy $S_i$ in detection model prediction (M-lc stands for low confidence, that is, high entropy)
    \item M-consistent and M-inconsistent: If low entropy $S_i$, and $M$ predicted label equals weak label then it is consistent (opposite predicted label and weak label, then inconsistent)
    \item S-unk: no social context signal, either its user's community is not dominantly reliable or unreliable/conspiracy, but a mixture; or the user is not clustered in any main community. 
    \item S-consistent and S-inconsistent: If the social context of a user account (its community label) is (in)consistent with the weak label of its tweet in $x_i$ (as described earlier).
\end{itemize}
The \textbf{objective} of the procedure is to minimize human relabel queries, and incorporate high confidence signals from both detection model $M$ and social context $S$ to ultimately remove or correct as many inaccurate weak labels, keeping as many correctly weak labeled instances. If the signals reinforce each other, the procedure can more confidently take an action without human label querying (or removal/discarding of the instance). Given the state, the appropriate action is selected by the procedure Alg.~\ref{alg:labelrefinement}.

\begin{algorithm}[t]  
  \caption{Label Refinement Procedure}
  \label{alg:labelrefinement}  
  \begin{algorithmic}[1]  
    \REQUIRE  
      Dataset instance $x_i$, weak label $y_i$, and detection model $M$, and social context $S$
    \ENSURE
      Action: RETAIN, FLIP, QUERY label
    \IF{M-consistent and S-consistent}
        \STATE RETAIN with weak label // reinforced signals from M and S
    \ELSIF{M-inconsistent and S-inconsistent}
        \STATE FLIP weak label // reinforced signals from M and S
    \ELSIF{(M-consistent and S-inconsistent) or (M-inconsistent and S-consistent)}
        \STATE QUERY label // contrasting signals from M and S
    \ELSIF{M-consistent and S-unk}
        \STATE RETAIN with weak label  // only signals from M
    \ELSIF{M-inconsistent and S-unk}
        \STATE FLIP weak label // only signals from M
    \ELSE
        \STATE QUERY label // detection model high entropy filtering
    \ENDIF
  \end{algorithmic}  
\end{algorithm}

\paragraph{Fine-grained semi-supervised classification} The dataset is refined based on retaining weak labels, model based relabeling, and human relabeling or removal of the instance. The retained and refined instances form the output constructed dataset with the associated model confidence in its label. The fine-grained labels are obtained by the human labeling but only on selected instances \textbf{false, unproven, mixture, mostly false, mostly true, true, and debunk}. For the remaining instances, we can use a \emph{semi-supervised classification} setup \cite{kipf2016semi,xie2020unsupervised} to obtain fine-grained distinctions. The few obtained human labeled instances become class prototypes to separate the rest into the seven classes. The distinctions can be very nuanced with varying degrees of truth, and difficult for a model to distinguish very accurately, so we provide these as auxiliary outputs. 

\begin{table*}[t]
\centering
\caption{Results on classification performance on test set from detection model with label refinement proposed approach for misinformation dataset construction on COVID-19 vaccines. Metrics: AP (average precision), AUC (ROC), F1 and Macro F1.}
\label{tab:misinformation_labeling_prelim_results}
\renewcommand*{\arraystretch}{0.7}
\begin{tabular}{@{}lllll@{}}
\toprule
Experiment                                 & AP             & AUC            & F1            & Macro F1        \\ \midrule
Weak labels                                & 0.722 $\pm$ 0.03 & 0.876 $\pm$ 0.01 & 0.774 $\pm$ 0.02 & 0.812 $\pm$ 0.01 \\
Self-training (iteration 1)                & 0.768 $\pm$ 0.01 & 0.888 $\pm$ 0.0  & 0.812 $\pm$ 0.01 & 0.842 $\pm$ 0.01 \\
Self-training (iteration 2)                & 0.775 $\pm$ 0.02 & 0.891 $\pm$ 0.0  & 0.811 $\pm$ 0.01 & 0.842 $\pm$ 0.01 \\
Social-context only                        & 0.764 $\pm$ 0.02 & 0.891 $\pm$ 0.01 & 0.810 $\pm$ 0.01 & 0.837 $\pm$ 0.01 \\
Social+Detection model                     & 0.785 $\pm$ 0.02 & 0.895 $\pm$ 0.0  & 0.813 $\pm$ 0.01 & 0.842 $\pm$ 0.01 \\
Social+Detection (+label correction) & \textbf{0.800 $\pm$ 0.01} & \textbf{0.895 $\pm$ 0.0}  & \textbf{0.818 $\pm$ 0.01} & \textbf{0.845 $\pm$ 0.01} \\ \bottomrule
\end{tabular}
\end{table*}

\begin{table*}[t]
\caption{Results for noise detection in weak labels with label refinement proposed approach for misinformation dataset construction on COVID-19 vaccines. Evaluation metrics: Rec (noise recall), Prec (precision), Frac UQ (fraction of unwanted queries), F1 (F1 of detected noise in weak labels).}
\label{tab:misinformation_labeling_recall_results}
\renewcommand*{\arraystretch}{0.5}
\resizebox{\textwidth}{!}{
\begin{tabular}{lllllllll}
\toprule
& \multicolumn{4}{c}{Test set} & \multicolumn{4}{c}{Validation set} \\
\midrule
\textbf{Experiment}  
& \textbf{Rec} & \textbf{Prec} & \textbf{Frac UQ} & \textbf{F1} & \textbf{Rec} & \textbf{Prec} & \textbf{Frac UQ} & \textbf{F1} \\
\midrule
Naive & 1.0  & 0.1719 & 1.0  & 0.2934 & 1.0  & 0.1533 & 1.0 & 0.2658 \\
Self-training (iteration 2) & 0.5682 & 0.3846  & 0.1887  & 0.4587 & 0.6522 & 0.4054  & 0.1732  & 0.5000 \\
Social-context only  & 0.4545 & 0.4444 & 0.1179 & 0.4494 & 0.4348 & 0.3704 & 0.1339 & 0.4000 \\
Social+Detection model & 0.8409 & 0.3978 & 0.2642 & 0.5401 & 0.7826 & 0.3673 & 0.2441 & 0.5000 \\
Social+Detection (+label flipping) & 0.8409  & 0.3978 & 0.2406 & 0.5401 & 0.7826  & 0.3673 & 0.2126 & 0.5000   \\  
\bottomrule
\end{tabular}
}
\end{table*}

\begin{table*}[t]
\caption{Fine-grained classification from human labeled class prototypes to remaining examples in the dataset.}
\label{tab:semisup}
\resizebox{\textwidth}{!}{
\begin{tabular}{lllllllllll}
\toprule
\multicolumn{1}{c}{\textbf{Expt (5-fold)}} & \multicolumn{1}{c}{\textbf{Macro F1}} & \multicolumn{1}{c}{\textbf{Weighted F1}} & \multicolumn{1}{c}{\textbf{F1 (debunk)}} & \multicolumn{1}{c}{\textbf{
(true)}} & \multicolumn{1}{c}{\textbf{(mostly\_true)}} & \multicolumn{1}{c}{\textbf{(mixture)}} & \multicolumn{1}{c}{\textbf{(mostly\_false)}} & \multicolumn{1}{c}{\textbf{(false)}} & \multicolumn{1}{c}{\textbf{(unproven)}} & \multicolumn{1}{c}{\textbf{Acc}} \\
\midrule
Random & 0.113 +/- 0.037 & 0.201 +/- 0.067 & 0.071 +/- 0.089 & 0.254 +/- 0.111 & 0.042 +/- 0.084 & 0.16 +/- 0.101 & 0.125 +/- 0.125 & 0.097 +/- 0.058 & 0.044 +/- 0.089 & 0.164 +/- 0.043 \\
Majority & 0.105 +/- 0.007 & 0.428 +/- 0.078 & 0.0 +/- 0.0 & 0.733 +/- 0.049 & 0.0 +/- 0.0 & 0.0 +/- 0.0 & 0.0 +/- 0.0 & 0.0 +/- 0.0 & 0.0 +/- 0.0 & 0.581 +/- 0.063 \\
Unweighted MLP & 0.219 +/- 0.022 & 0.552 +/- 0.073 & 0.04 +/- 0.08 & 0.757 +/- 0.045 & 0.0 +/- 0.0 & 0.257 +/- 0.074 & 0.0 +/- 0.0 & 0.483 +/- 0.064 & 0.0 +/- 0.0 & 0.595 +/- 0.054 \\
Weighted MLP & 0.266 +/- 0.055 & 0.57 +/- 0.077 & 0.235 +/- 0.145 & 0.735 +/- 0.05 & 0.033 +/- 0.067 & 0.338 +/- 0.165 & 0.0 +/- 0.0 & 0.52 +/- 0.1 & 0.0 +/- 0.0 & 0.546 +/- 0.061 \\
\bottomrule
\end{tabular}
}
\end{table*}

\section{Experiments}
\label{sec:experiments}

We study the proposed approach for constructing a \textbf{large-scale public misinformation dataset} on \textbf{COVID-19 vaccines}. We use iterative self-training of the misinformation detection model CSI  \cite{ruchansky2017csi} trained first on the initial weakly-labeled cascades. We use low-quality news sources for weak labels on the collected Twitter dataset, compiled from fact-checking resources as described earlier in Section~\ref{sec:data}. We have a total of 14.6k tweet cascades with roughly 10,377 weakly labeled as reliable and 4,267 weakly labeled as unreliable/conspiracy. With this setting, we experiment with the proposed framework for large-scale misinformation dataset construction from weak news-source labels.

\subsection{Evaluation} The evaluation \textbf{test set} of tweet cascades contains \textbf{256 tweets} with ground-truth fact-checked claim based labels obtained by searching for tweets related to Snopes fact-checks and AP news/NPR news on COVID-19 vaccines and labeling from the 7 fine-grained labels according to the annotation scheme and fact-checked claims. For experiments, a human-labeled \textbf{validation set} of \textbf{150 tweets}, based on the annotation scheme and guidelines, is also constructed and held-out from the \textbf{14.6k tweet cascades} (as described in Sec~\ref{sec:correlation_sets}).

\subsubsection{Evaluation tasks.} We cannot directly measure the quality of the constructed misinformation dataset, since we cannot obtain ground-truth fact-checker (e.g. Snopes) labels on all 14.6k tweet cascades. We instead evaluate on the fact-checked claim based test subset of 256 tweet cascades using the following evaluation metrics (i) \textbf{Misinformation detection performance} on test set. Label quality in the dataset should be positively correlated with misinformation detection accuracy on ground-truth labeled data. (ii) \textbf{Label correction accuracy} on validation and test sets and (iii) \# of wasted queries generated in the label refinement procedure, to measure human \textbf{resources that are inefficiently utilized}.

\subsubsection{Evaluation metrics.} (i) The baselines and proposed experiments are evaluated for \textbf{misinformation detection (classification) performance} on the ground-truth test set of 256 tweet cascades. The detection model performance is averaged over 5 random seeds. The evaluation includes the classification metrics namely, Area under the precision-recall curve (AP) and Area under the ROC curve (AUC) and F1 and Macro-F1 for the misinformation classification. It measures how well the refined/constructed dataset from baselines or the proposed approach work in separating the misinformation from reliable information tweet cascades.

(ii) The baselines and proposed experiments are evaluated for \textbf{label correction accuracy} on the ground-truth test set and validation set. We have the initial weak labels and correct misinformation labels for the test and validation set. Therefore, we can measure the recall (Rec), precision (Prec), and F1 of the noise in the weak labels (i.e., weak label and ground-truth fact-checked label are not aligned). The instances detected as noise by the methods are ones selected for FLIP or QUERY (REMOVE) actions (as it is predicted by the method as having a possibly mislabeled weak label). Recall is the fraction of actual noise in weak labels that are correctly detected by the methods, and Precision measures the correctly recalled noise in weak labels out of all instances detected as noise by the methods. F1 is the harmonic mean of precision and recall.

(iii) We additionally propose a metric to also measure how \textbf{efficiently the resources are utilized} by the baselines and proposed methods. We define Frac UQ (fraction of correctly weak-labeled instances that are assigned QUERY (REMOVE) action for human relabeling (removal), i.e. unwanted or wasted queries) as follows,
\begin{equation}
    \textrm{Frac UQ} = \frac{|\textrm{ (QUERY action assigned) } \& \textrm{ (correct weak label) }|}{|\textrm{correct weak label}|}
\end{equation}
The \# of instances with correct weak labels assigned QUERY (REMOVE) action is the numerator, measuring human resource wastage. \emph{Lower value of Frac UQ is better}, while maintaining high noise recall. 



\subsection{Results} 

\textbf{Misinformation detection performance.} In \textbf{Table~\ref{tab:misinformation_labeling_prelim_results}} we provide results of the proposed framework to construct misinformation datasets from weak labels. We trained the CSI \cite{ruchansky2017csi} misinformation detection model on weak labels from news source credibility to classify misinformation (unreliable/conspiracy) tweets from reliable information tweets, as a baseline. The held-out validation set is used by the detection model for early stopping in model optimization, and for calculating the threshold for detection based on AUC curve, trading off sensitivity and specificity on the validation set. The reported results are on the held-out ground-truth test set of fact-checked based labels. The same setting is used in all experiments.
 
The removal (entropy filtering) guided by the detection model (self-training iteration 1 and 2) improves the classification on the ground-truth test set, indicative of higher label quality in the retained tweets. After 2 iterations, the improvement was insignificant.

Further, incorporating social context modeling, we first evaluate Social-context only, wherein the tweets with labels opposite to their community label (dominantly reliable or dominantly misinformation) are surfaced as to be queried or removed. We find that combining the social context modeling and detection model guidance is more informative about possible mislabeling (tweets to be removed) in the weak labels (Social$+$Detection model). Finally, Social$+$Detection model ($+$label correction) is used to correct the labels that the two signals suggest should be oppositely labeled, and remove ones that the model is unsure about either from the detection model or social context (i.e., using the label refinement procedure in Alg 1). We find results in model-guided label refinement for construction of misinformation datasets is effective and significantly improves both recall in misinformation detection, (since the misinformation examples are fewer in the imbalanced data), and the precision of detected misinformation, and other metrics separating the two classes of misinformation and reliable information.

\textbf{Label correction accuracy and resource efficiency of label refinement.} In \textbf{Table~\ref{tab:misinformation_labeling_recall_results}}, we provide the results of performance on label correction using the signals from the detection model and/or social context. Naive baseline is trivially set to assume all weak labels are mislabeled, and QUERY all of them. Therefore, all correctly weak-labeled instances are Queried with worst resource utilization of 1, and low precision, F1 scores. For the removal (entropy filtering) guided by the detection model (self-training iteration 2) has roughly 56\% recall in inaccurate weak labels, with reasonable precision and low Frac UQ. It is similarly the case for Social-context only method. 

With the proposed approach (i.e., using the label refinement procedure in Alg 1) combining the social context modeling and detection model guidance is more informative about possible mislabeling in the weak labels (Social+Detection model) and we see a massive increase in recall on combining the two signals, resulting in 0.84 recall. This might suggest that the signals provided by detection model entropy filtering and social context are \textbf{complementary} to each other, and jointly inform label refinement in the proposed algorithm most effectively. With label FLIP action  included (Social+Detection (+label flipping), the difference is only in Frac UQ, where now some of the detected noise will be directly selected for FLIP action instead of for removal (or query), minimizing the wasted queries, if the FLIP was assigned to actual noisy instances.

These evaluation metrics suggest how well the proposed method works at constructing high-quality misinformation labels, with the least cost incurred in terms of the human labeling resources, or mistakes in identification of possible incorrect weak labels.

\begin{figure}[t]
    \centering
    \begin{subfigure}{0.49\columnwidth}
    \includegraphics[scale=0.4]{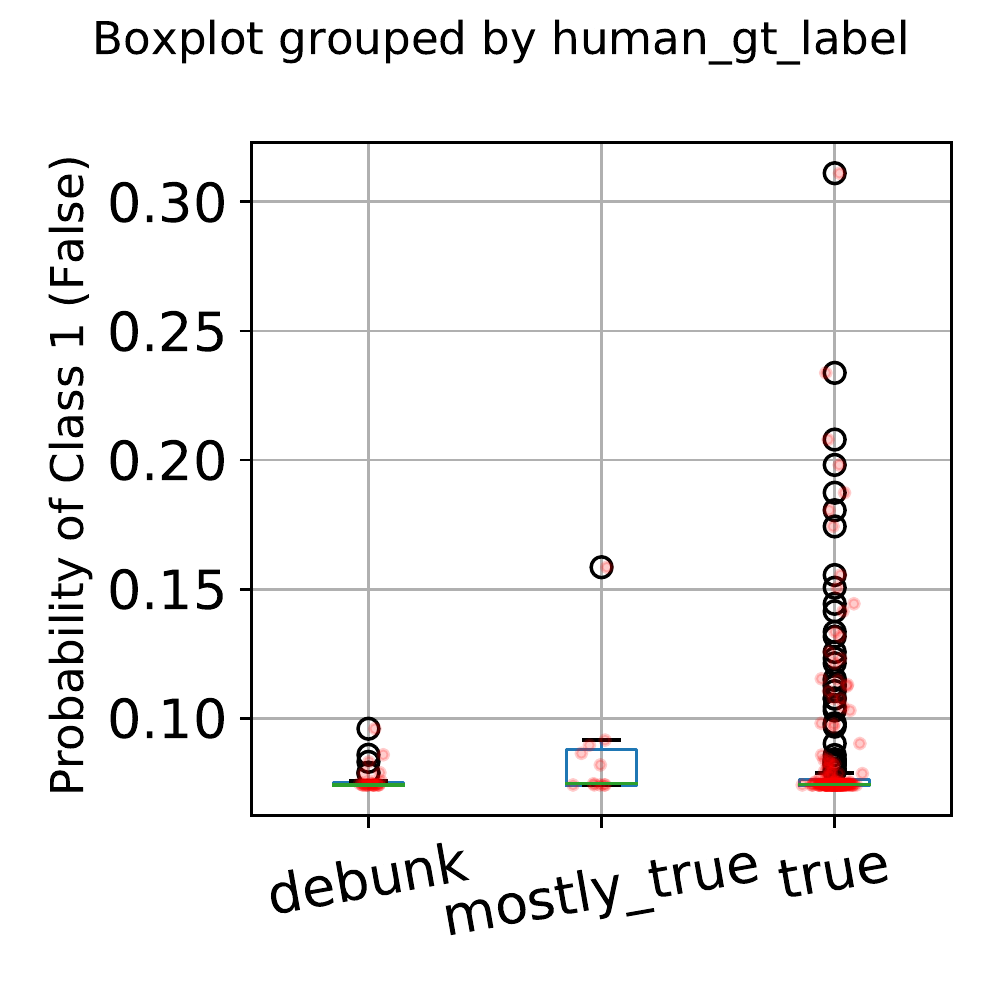}
    \end{subfigure}
    ~
    \begin{subfigure}{0.49\columnwidth}
    \includegraphics[scale=0.4]{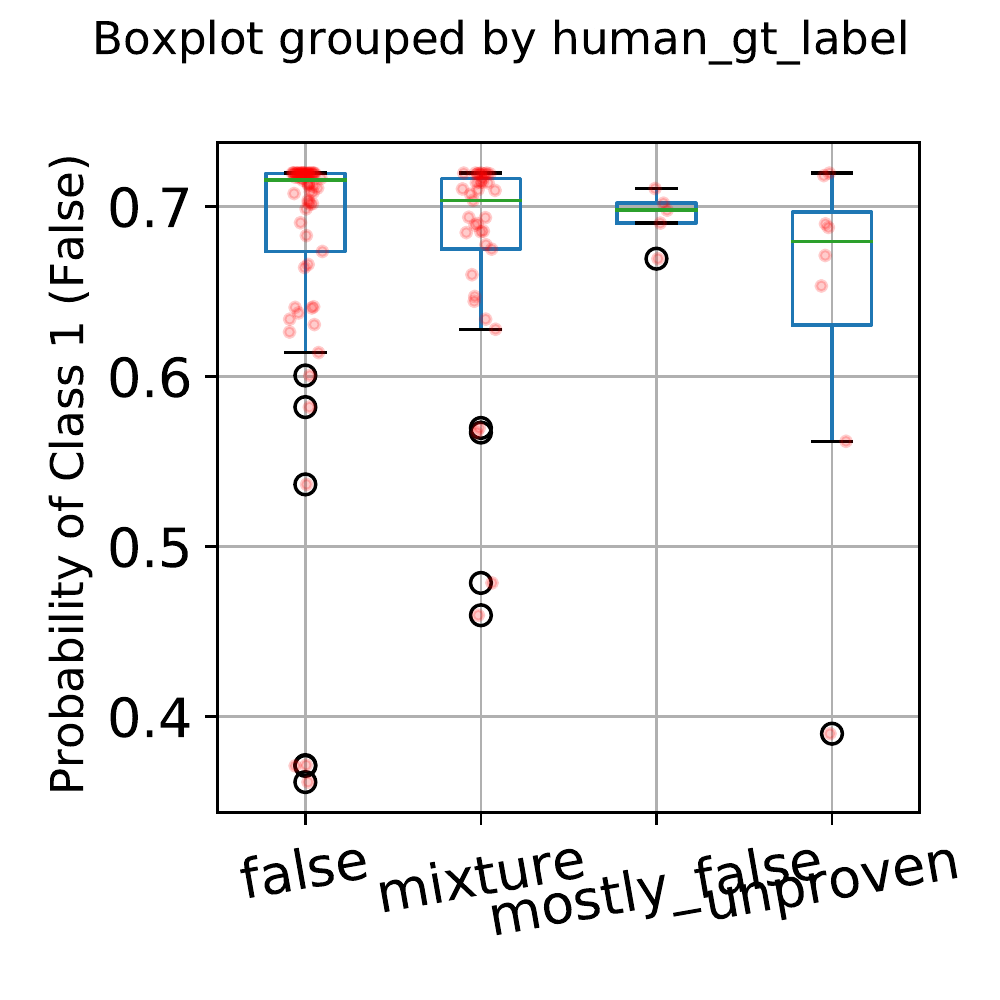}
    \end{subfigure}
    \caption{Predicted probabilities from detection model after label refinement:  correlation with true labels on ground-truth fact-checked human-labeled test and validation set.}
    \label{fig:correlation_prob_fake}
\end{figure}

\textbf{Constructed misinformation dataset.}  
In the constructed misinformation dataset derived from weak labels with the proposed method, in Fig.~\ref{fig:correlation_prob_fake}, we examine scatter plot of instances on the predicted probability of misinformation from the detection model, which as we see is correlated with the fine-grained human labels available on the validation and test set, capturing the varying degree of truth. In Table~\ref{tab:semisup}, we show the fine-grained classification from human labeled class prototypes on remaining examples in the dataset, using 5-fold cross validation on stratified splits of the validation plus test set for evaluation. For classification, we additionally labeled 400 instances to include as human-labeled class prototypes. We used extracted representations of tweet cascades from the CSI detection model used here, to train an MLP. With class-weighting, the fine-grained classifier has 0.57 weighted F1 distinguishing over the 7 nuanced label categories which is a difficult task. 

\section{Discussion and Conclusions}

The proposed label refinement approach is effective at constructing large-scale datasets from weak labels with high recall of inaccurate weak labels when incorporating social context jointly with entropy filtering. We provide discussions of the proposed approach and three potential, concrete future research directions: (1) The weak labels are collected from news-source credibility, so posts with references to news sources form the basis of the dataset. The dataset might be more centered on  news-worthy contents, which is one limitation of the approach. Other sources to augment weak labels could be considered in the future. Also, the annotation scheme label categories are general, but effective ways to construct expert examples as annotator guidelines in new domains could be explored. 

(2) The proposed approach models \emph{instance credibility} and \emph{user credibility} through entropy filtering and social context modeling. The label refinement procedure could be provided additional signals of news-source credibility by modeling each news source separately (since each might cause different noise rates, and unreliable sources are more mixed than conspiracy sources, as we observed). (3) The fine-grained classification is a difficult task and future research directions could explore model-guided selection of instances for human labeling, to act as the most effective class prototypes, with richer semi-supervised fine-grained classification techniques. 

To conclude, the proposed label refinement with social context modeling is a useful, new approach for constructing misinformation datasets, in a timely and scalable way for new or evolving domains.






\begin{acks} 
This work is supported by NSF Research Grant (CCF-1837131) and DARPA (HR001121C0169 and W911NF-17-C-0094). Views and conclusions are of the authors and should not be interpreted as representing the social policies of the funding agency, or the U.S. Government. We thank Feng Pan and Cindy Lin for helping with annotation guidelines and labeling efforts. 
\end{acks}

\bibliographystyle{ACM-Reference-Format}
\bibliography{main}


\begin{thebibliography}{35}


\ifx \showCODEN    \undefined \def \showCODEN     #1{\unskip}     \fi
\ifx \showDOI      \undefined \def \showDOI       #1{#1}\fi
\ifx \showISBNx    \undefined \def \showISBNx     #1{\unskip}     \fi
\ifx \showISBNxiii \undefined \def \showISBNxiii  #1{\unskip}     \fi
\ifx \showISSN     \undefined \def \showISSN      #1{\unskip}     \fi
\ifx \showLCCN     \undefined \def \showLCCN      #1{\unskip}     \fi
\ifx \shownote     \undefined \def \shownote      #1{#1}          \fi
\ifx \showarticletitle \undefined \def \showarticletitle #1{#1}   \fi
\ifx \showURL      \undefined \def \showURL       {\relax}        \fi
\providecommand\bibfield[2]{#2}
\providecommand\bibinfo[2]{#2}
\providecommand\natexlab[1]{#1}
\providecommand\showeprint[2][]{arXiv:#2}

\bibitem[\protect\citeauthoryear{Arazo, Ortego, Albert, O’Connor, and
  Mcguinness}{Arazo et~al\mbox{.}}{2019}]%
        {arazo2019unsupervised}
\bibfield{author}{\bibinfo{person}{Eric Arazo}, \bibinfo{person}{Diego Ortego},
  \bibinfo{person}{Paul Albert}, \bibinfo{person}{Noel O’Connor}, {and}
  \bibinfo{person}{Kevin Mcguinness}.} \bibinfo{year}{2019}\natexlab{}.
\newblock \showarticletitle{Unsupervised Label Noise Modeling and Loss
  Correction}. In \bibinfo{booktitle}{\emph{International Conference on Machine
  Learning}}. \bibinfo{pages}{312--321}.
\newblock


\bibitem[\protect\citeauthoryear{Blondel, Guillaume, Lambiotte, and
  Lefebvre}{Blondel et~al\mbox{.}}{2008}]%
        {blondel2008fast}
\bibfield{author}{\bibinfo{person}{Vincent~D Blondel},
  \bibinfo{person}{Jean-Loup Guillaume}, \bibinfo{person}{Renaud Lambiotte},
  {and} \bibinfo{person}{Etienne Lefebvre}.} \bibinfo{year}{2008}\natexlab{}.
\newblock \showarticletitle{Fast unfolding of communities in large networks}.
\newblock \bibinfo{journal}{\emph{Journal of statistical mechanics: theory and
  experiment}} \bibinfo{volume}{2008}, \bibinfo{number}{10}
  (\bibinfo{year}{2008}), \bibinfo{pages}{P10008}.
\newblock


\bibitem[\protect\citeauthoryear{Bond}{Bond}{2021}]%
        {nprbirdwatch}
\bibfield{author}{\bibinfo{person}{Shannon Bond}.}
  \bibinfo{year}{2021}\natexlab{}.
\newblock \bibinfo{booktitle}{\emph{Twitter's 'Birdwatch' Aims to Crowdsource
  Fight Against Misinformation}}.
\newblock
\urldef\tempurl%
\url{https://www.npr.org/2021/02/10/965839888/twitters-birdwatch-aims-to-crowdsource-fight-against-misinformation}
\showURL{%
Retrieved 2021 from \tempurl}


\bibitem[\protect\citeauthoryear{Bozarth, Saraf, and Budak}{Bozarth
  et~al\mbox{.}}{2020}]%
        {Bozarth_Saraf_Budak_2020}
\bibfield{author}{\bibinfo{person}{Lia Bozarth}, \bibinfo{person}{Aparajita
  Saraf}, {and} \bibinfo{person}{Ceren Budak}.}
  \bibinfo{year}{2020}\natexlab{}.
\newblock \showarticletitle{Higher Ground? How Groundtruth Labeling Impacts Our
  Understanding of Fake News about the 2016 U.S. Presidential Nominees}.
\newblock \bibinfo{journal}{\emph{Proceedings of the International AAAI
  Conference on Web and Social Media}} \bibinfo{volume}{14},
  \bibinfo{number}{1} (\bibinfo{date}{May} \bibinfo{year}{2020}),
  \bibinfo{pages}{48--59}.
\newblock
\urldef\tempurl%
\url{https://ojs.aaai.org/index.php/ICWSM/article/view/7278}
\showURL{%
\tempurl}


\bibitem[\protect\citeauthoryear{Dai, Sun, and Wang}{Dai et~al\mbox{.}}{2020}]%
        {dai2020ginger}
\bibfield{author}{\bibinfo{person}{Enyan Dai}, \bibinfo{person}{Yiwei Sun},
  {and} \bibinfo{person}{Suhang Wang}.} \bibinfo{year}{2020}\natexlab{}.
\newblock \showarticletitle{Ginger Cannot Cure Cancer: Battling Fake Health
  News with a Comprehensive Data Repository}. In
  \bibinfo{booktitle}{\emph{Proceedings of the International AAAI Conference on
  Web and Social Media}}, Vol.~\bibinfo{volume}{14}. \bibinfo{pages}{853--862}.
\newblock


\bibitem[\protect\citeauthoryear{Fefilatyev, Shreve, Kramer, Hall, Goldgof,
  Kasturi, Daly, Remsen, and Bunke}{Fefilatyev et~al\mbox{.}}{2012}]%
        {fefilatyev2012label}
\bibfield{author}{\bibinfo{person}{Sergiy Fefilatyev}, \bibinfo{person}{Matthew
  Shreve}, \bibinfo{person}{Kurt Kramer}, \bibinfo{person}{Lawrence Hall},
  \bibinfo{person}{Dmitry Goldgof}, \bibinfo{person}{Rangachar Kasturi},
  \bibinfo{person}{Kendra Daly}, \bibinfo{person}{Andrew Remsen}, {and}
  \bibinfo{person}{Horst Bunke}.} \bibinfo{year}{2012}\natexlab{}.
\newblock \showarticletitle{Label-noise reduction with support vector
  machines}. In \bibinfo{booktitle}{\emph{Proceedings of the 21st International
  Conference on Pattern Recognition (ICPR2012)}}. IEEE,
  \bibinfo{pages}{3504--3508}.
\newblock


\bibitem[\protect\citeauthoryear{Ferrara, Chang, Chen, Muric, and
  Patel}{Ferrara et~al\mbox{.}}{2020}]%
        {ferrara2020characterizing}
\bibfield{author}{\bibinfo{person}{Emilio Ferrara}, \bibinfo{person}{Herbert
  Chang}, \bibinfo{person}{Emily Chen}, \bibinfo{person}{Goran Muric}, {and}
  \bibinfo{person}{Jaimin Patel}.} \bibinfo{year}{2020}\natexlab{}.
\newblock \showarticletitle{Characterizing social media manipulation in the
  2020 US presidential election}.
\newblock \bibinfo{journal}{\emph{First Monday}} (\bibinfo{year}{2020}).
\newblock


\bibitem[\protect\citeauthoryear{Fr{\'e}nay and Verleysen}{Fr{\'e}nay and
  Verleysen}{2013}]%
        {frenay2013classification}
\bibfield{author}{\bibinfo{person}{Beno{\^\i}t Fr{\'e}nay} {and}
  \bibinfo{person}{Michel Verleysen}.} \bibinfo{year}{2013}\natexlab{}.
\newblock \showarticletitle{Classification in the presence of label noise: a
  survey}.
\newblock \bibinfo{journal}{\emph{IEEE transactions on neural networks and
  learning systems}} \bibinfo{volume}{25}, \bibinfo{number}{5}
  (\bibinfo{year}{2013}), \bibinfo{pages}{845--869}.
\newblock


\bibitem[\protect\citeauthoryear{Garimella, Morales, Gionis, and
  Mathioudakis}{Garimella et~al\mbox{.}}{2018}]%
        {garimella2018quantifying}
\bibfield{author}{\bibinfo{person}{Kiran Garimella}, \bibinfo{person}{Gianmarco
  De~Francisci Morales}, \bibinfo{person}{Aristides Gionis}, {and}
  \bibinfo{person}{Michael Mathioudakis}.} \bibinfo{year}{2018}\natexlab{}.
\newblock \showarticletitle{Quantifying controversy on social media}.
\newblock \bibinfo{journal}{\emph{ACM Transactions on Social Computing}}
  \bibinfo{volume}{1}, \bibinfo{number}{1} (\bibinfo{year}{2018}),
  \bibinfo{pages}{1--27}.
\newblock


\bibitem[\protect\citeauthoryear{Kipf and Welling}{Kipf and Welling}{2016}]%
        {kipf2016semi}
\bibfield{author}{\bibinfo{person}{Thomas~N Kipf} {and} \bibinfo{person}{Max
  Welling}.} \bibinfo{year}{2016}\natexlab{}.
\newblock \showarticletitle{Semi-supervised classification with graph
  convolutional networks}.
\newblock \bibinfo{journal}{\emph{arXiv preprint arXiv:1609.02907}}
  (\bibinfo{year}{2016}).
\newblock


\bibitem[\protect\citeauthoryear{Kremer, Sha, and Igel}{Kremer
  et~al\mbox{.}}{2018}]%
        {kremer2018robust}
\bibfield{author}{\bibinfo{person}{Jan Kremer}, \bibinfo{person}{Fei Sha},
  {and} \bibinfo{person}{Christian Igel}.} \bibinfo{year}{2018}\natexlab{}.
\newblock \showarticletitle{Robust active label correction}. In
  \bibinfo{booktitle}{\emph{International conference on artificial intelligence
  and statistics}}. PMLR, \bibinfo{pages}{308--316}.
\newblock


\bibitem[\protect\citeauthoryear{Kwon, Cha, and Jung}{Kwon
  et~al\mbox{.}}{2016}]%
        {kwondata}
\bibfield{author}{\bibinfo{person}{Sejeong Kwon}, \bibinfo{person}{Meeyoung
  Cha}, {and} \bibinfo{person}{Kyomin Jung}.} \bibinfo{year}{2016}\natexlab{}.
\newblock \showarticletitle{{Rumor detection over varying time windows.}}. In
  \bibinfo{booktitle}{\emph{Harvard Dataverse}}. \bibinfo{publisher}{Harvard
  Dataverse}.
\newblock
\urldef\tempurl%
\url{https://doi.org/10.7910/DVN/BFGAVZ}
\showDOI{\tempurl}


\bibitem[\protect\citeauthoryear{Lewandowsky, Cook, Schmid, Holford, Finn,
  Leask, Thomson, Lombardi, Al-Rawi, Amazeen, et~al\mbox{.}}{Lewandowsky
  et~al\mbox{.}}{2021}]%
        {lewandowsky2021covid}
\bibfield{author}{\bibinfo{person}{Stephan Lewandowsky}, \bibinfo{person}{John
  Cook}, \bibinfo{person}{Philipp Schmid}, \bibinfo{person}{Dawn~Liu Holford},
  \bibinfo{person}{Adam Finn}, \bibinfo{person}{Julie Leask},
  \bibinfo{person}{Angus Thomson}, \bibinfo{person}{Doug Lombardi},
  \bibinfo{person}{Ahmed~K Al-Rawi}, \bibinfo{person}{Michelle~A Amazeen},
  {et~al\mbox{.}}} \bibinfo{year}{2021}\natexlab{}.
\newblock \bibinfo{title}{The COVID-19 Vaccine Communication Handbook. A
  practical guide for improving vaccine communication and fighting
  misinformation}.
\newblock
\newblock


\bibitem[\protect\citeauthoryear{Ma, Gao, Mitra, Kwon, Jansen, Wong, and
  Cha}{Ma et~al\mbox{.}}{2016}]%
        {ma2016detecting}
\bibfield{author}{\bibinfo{person}{Jing Ma}, \bibinfo{person}{Wei Gao},
  \bibinfo{person}{Prasenjit Mitra}, \bibinfo{person}{Sejeong Kwon},
  \bibinfo{person}{Bernard~J Jansen}, \bibinfo{person}{Kam-Fai Wong}, {and}
  \bibinfo{person}{Meeyoung Cha}.} \bibinfo{year}{2016}\natexlab{}.
\newblock \showarticletitle{Detecting Rumors from Microblogs with Recurrent
  Neural Networks.}. In \bibinfo{booktitle}{\emph{IJCAI}}.
  \bibinfo{pages}{3818--3824}.
\newblock


\bibitem[\protect\citeauthoryear{Memon and Carley}{Memon and Carley}{2020}]%
        {memon2020characterizing}
\bibfield{author}{\bibinfo{person}{Shahan~Ali Memon} {and}
  \bibinfo{person}{Kathleen~M Carley}.} \bibinfo{year}{2020}\natexlab{}.
\newblock \showarticletitle{Characterizing COVID-19 misinformation communities
  using a novel twitter dataset}. In \bibinfo{booktitle}{\emph{CEUR Workshop
  Proceedings}}, Vol.~\bibinfo{volume}{2699}.
\newblock


\bibitem[\protect\citeauthoryear{Popat, Mukherjee, Strötgen, and Weikum}{Popat
  et~al\mbox{.}}{2017}]%
        {popat_where_2017}
\bibfield{author}{\bibinfo{person}{Kashyap Popat}, \bibinfo{person}{Subhabrata
  Mukherjee}, \bibinfo{person}{Jannik Strötgen}, {and}
  \bibinfo{person}{Gerhard Weikum}.} \bibinfo{year}{2017}\natexlab{}.
\newblock \showarticletitle{Where the {Truth} {Lies}: {Explaining} the
  {Credibility} of {Emerging} {Claims} on the {Web} and {Social} {Media}}. In
  \bibinfo{booktitle}{\emph{Proceedings of the 26th {International}
  {Conference} on {World} {Wide} {Web} {Companion} - {WWW} '17 {Companion}}}.
  \bibinfo{publisher}{ACM Press}, \bibinfo{address}{Perth, Australia},
  \bibinfo{pages}{1003--1012}.
\newblock
\showISBNx{978-1-4503-4914-7}
\urldef\tempurl%
\url{https://doi.org/10.1145/3041021.3055133}
\showDOI{\tempurl}


\bibitem[\protect\citeauthoryear{Qian, Gong, Sharma, and Liu}{Qian
  et~al\mbox{.}}{2018}]%
        {qian2018neural}
\bibfield{author}{\bibinfo{person}{Feng Qian}, \bibinfo{person}{Chengyue Gong},
  \bibinfo{person}{Karishma Sharma}, {and} \bibinfo{person}{Yan Liu}.}
  \bibinfo{year}{2018}\natexlab{}.
\newblock \showarticletitle{Neural User Response Generator: Fake News Detection
  with Collective User Intelligence.}. In \bibinfo{booktitle}{\emph{IJCAI}},
  Vol.~\bibinfo{volume}{3834}. \bibinfo{pages}{3840}.
\newblock


\bibitem[\protect\citeauthoryear{Ruchansky, Seo, and Liu}{Ruchansky
  et~al\mbox{.}}{2017}]%
        {ruchansky2017csi}
\bibfield{author}{\bibinfo{person}{Natali Ruchansky}, \bibinfo{person}{Sungyong
  Seo}, {and} \bibinfo{person}{Yan Liu}.} \bibinfo{year}{2017}\natexlab{}.
\newblock \showarticletitle{CSI: A Hybrid Deep Model for Fake News Detection}.
  In \bibinfo{booktitle}{\emph{Proceedings of the 2017 ACM on Conference on
  Information and Knowledge Management}}. ACM, \bibinfo{pages}{797--806}.
\newblock


\bibitem[\protect\citeauthoryear{Salem, Al~Feel, Elbassuoni, Jaber, and
  Farah}{Salem et~al\mbox{.}}{2019}]%
        {salem2019fa}
\bibfield{author}{\bibinfo{person}{Fatima K~Abu Salem}, \bibinfo{person}{Roaa
  Al~Feel}, \bibinfo{person}{Shady Elbassuoni}, \bibinfo{person}{Mohamad
  Jaber}, {and} \bibinfo{person}{May Farah}.} \bibinfo{year}{2019}\natexlab{}.
\newblock \showarticletitle{FA-KES: a fake news dataset around the Syrian war}.
  In \bibinfo{booktitle}{\emph{Proceedings of the International AAAI Conference
  on Web and Social Media}}, Vol.~\bibinfo{volume}{13}.
  \bibinfo{pages}{573--582}.
\newblock


\bibitem[\protect\citeauthoryear{Sharma, Donmez, Luo, Liu, and Yalniz}{Sharma
  et~al\mbox{.}}{2020a}]%
        {sharma2020noiserank}
\bibfield{author}{\bibinfo{person}{Karishma Sharma}, \bibinfo{person}{Pinar
  Donmez}, \bibinfo{person}{Enming Luo}, \bibinfo{person}{Yan Liu}, {and}
  \bibinfo{person}{I~Zeki Yalniz}.} \bibinfo{year}{2020}\natexlab{a}.
\newblock \showarticletitle{Noiserank: Unsupervised label noise reduction with
  dependence models}. In \bibinfo{booktitle}{\emph{Computer Vision--ECCV 2020:
  16th European Conference, Glasgow, UK, August 23--28, 2020, Proceedings, Part
  XXVII 16}}. Springer, \bibinfo{pages}{737--753}.
\newblock


\bibitem[\protect\citeauthoryear{Sharma, Ferrara, and Liu}{Sharma
  et~al\mbox{.}}{2022}]%
        {sharma2022election}
\bibfield{author}{\bibinfo{person}{Karishma Sharma}, \bibinfo{person}{Emilio
  Ferrara}, {and} \bibinfo{person}{Yan Liu}.} \bibinfo{year}{2022}\natexlab{}.
\newblock \showarticletitle{Characterizing Online Engagement with
  Disinformation and Conspiracies in the 2020 U.S. Presidential Election}. In
  \bibinfo{booktitle}{\emph{Proceedings of the International AAAI Conference on
  Web and Social Media}}.
\newblock


\bibitem[\protect\citeauthoryear{Sharma, Qian, Jiang, Ruchansky, Zhang, and
  Liu}{Sharma et~al\mbox{.}}{2019}]%
        {sharma2019combating}
\bibfield{author}{\bibinfo{person}{Karishma Sharma}, \bibinfo{person}{Feng
  Qian}, \bibinfo{person}{He Jiang}, \bibinfo{person}{Natali Ruchansky},
  \bibinfo{person}{Ming Zhang}, {and} \bibinfo{person}{Yan Liu}.}
  \bibinfo{year}{2019}\natexlab{}.
\newblock \showarticletitle{Combating fake news: A survey on identification and
  mitigation techniques}.
\newblock \bibinfo{journal}{\emph{ACM Transactions on Intelligent Systems and
  Technology (TIST)}} \bibinfo{volume}{10}, \bibinfo{number}{3}
  (\bibinfo{year}{2019}), \bibinfo{pages}{1--42}.
\newblock


\bibitem[\protect\citeauthoryear{Sharma, Seo, Meng, Rambhatla, and Liu}{Sharma
  et~al\mbox{.}}{2020b}]%
        {sharma2020covid}
\bibfield{author}{\bibinfo{person}{Karishma Sharma}, \bibinfo{person}{Sungyong
  Seo}, \bibinfo{person}{Chuizheng Meng}, \bibinfo{person}{Sirisha Rambhatla},
  {and} \bibinfo{person}{Yan Liu}.} \bibinfo{year}{2020}\natexlab{b}.
\newblock \showarticletitle{Covid-19 on social media: Analyzing misinformation
  in twitter conversations}.
\newblock \bibinfo{journal}{\emph{arXiv e-prints}} (\bibinfo{year}{2020}),
  \bibinfo{pages}{arXiv--2003}.
\newblock


\bibitem[\protect\citeauthoryear{Sharma, Zhang, Ferrara, and Liu}{Sharma
  et~al\mbox{.}}{2021b}]%
        {sharma2020identifying}
\bibfield{author}{\bibinfo{person}{Karishma Sharma}, \bibinfo{person}{Yizhou
  Zhang}, \bibinfo{person}{Emilio Ferrara}, {and} \bibinfo{person}{Yan Liu}.}
  \bibinfo{year}{2021}\natexlab{b}.
\newblock \showarticletitle{Identifying Coordinated Accounts on Social Media
  through Hidden Influence and Group Behaviours}. In
  \bibinfo{booktitle}{\emph{Proceedings of the 27th ACM SIGKDD Conference on
  Knowledge Discovery \& Data Mining}} (Virtual Event, Singapore)
  \emph{(\bibinfo{series}{KDD '21})}. \bibinfo{publisher}{Association for
  Computing Machinery}, \bibinfo{pages}{1441–1451}.
\newblock
\showISBNx{9781450383325}
\urldef\tempurl%
\url{https://doi.org/10.1145/3447548.3467391}
\showDOI{\tempurl}


\bibitem[\protect\citeauthoryear{Sharma, Zhang, and Liu}{Sharma
  et~al\mbox{.}}{2021a}]%
        {sharma2021covid19vaccine}
\bibfield{author}{\bibinfo{person}{Karishma Sharma}, \bibinfo{person}{Yizhou
  Zhang}, {and} \bibinfo{person}{Yan Liu}.} \bibinfo{year}{2021}\natexlab{a}.
\newblock \bibinfo{title}{COVID-19 Vaccines: Characterizing Misinformation
  Campaigns and Vaccine Hesitancy on Twitter}.
\newblock
\newblock
\showeprint[arxiv]{2106.08423}~[cs.SI]


\bibitem[\protect\citeauthoryear{Shu, Dumais, Awadallah, and Liu}{Shu
  et~al\mbox{.}}{2020a}]%
        {shu2020detecting}
\bibfield{author}{\bibinfo{person}{Kai Shu}, \bibinfo{person}{Susan Dumais},
  \bibinfo{person}{Ahmed~Hassan Awadallah}, {and} \bibinfo{person}{Huan Liu}.}
  \bibinfo{year}{2020}\natexlab{a}.
\newblock \showarticletitle{Detecting fake news with weak social supervision}.
\newblock \bibinfo{journal}{\emph{IEEE Intelligent Systems}}
  \bibinfo{volume}{36}, \bibinfo{number}{4} (\bibinfo{year}{2020}),
  \bibinfo{pages}{96--103}.
\newblock


\bibitem[\protect\citeauthoryear{Shu, Mahudeswaran, Wang, Lee, and Liu}{Shu
  et~al\mbox{.}}{2020b}]%
        {shu2020fakenewsnet}
\bibfield{author}{\bibinfo{person}{Kai Shu}, \bibinfo{person}{Deepak
  Mahudeswaran}, \bibinfo{person}{Suhang Wang}, \bibinfo{person}{Dongwon Lee},
  {and} \bibinfo{person}{Huan Liu}.} \bibinfo{year}{2020}\natexlab{b}.
\newblock \showarticletitle{FakeNewsNet: A Data Repository with News Content,
  Social Context, and Spatiotemporal Information for Studying Fake News on
  Social Media}.
\newblock \bibinfo{journal}{\emph{Big Data}} \bibinfo{volume}{8},
  \bibinfo{number}{3} (\bibinfo{year}{2020}), \bibinfo{pages}{171--188}.
\newblock


\bibitem[\protect\citeauthoryear{Singh, Bansal, Bode, Budak, Chi, Kawintiranon,
  Padden, Vanarsdall, Vraga, and Wang}{Singh et~al\mbox{.}}{2020}]%
        {singh2020first}
\bibfield{author}{\bibinfo{person}{Lisa Singh}, \bibinfo{person}{Shweta
  Bansal}, \bibinfo{person}{Leticia Bode}, \bibinfo{person}{Ceren Budak},
  \bibinfo{person}{Guangqing Chi}, \bibinfo{person}{Kornraphop Kawintiranon},
  \bibinfo{person}{Colton Padden}, \bibinfo{person}{Rebecca Vanarsdall},
  \bibinfo{person}{Emily Vraga}, {and} \bibinfo{person}{Yanchen Wang}.}
  \bibinfo{year}{2020}\natexlab{}.
\newblock \showarticletitle{A first look at COVID-19 information and
  misinformation sharing on Twitter}.
\newblock \bibinfo{journal}{\emph{arXiv preprint arXiv:2003.13907}}
  (\bibinfo{year}{2020}).
\newblock


\bibitem[\protect\citeauthoryear{Song, Kim, Park, Shin, and Lee}{Song
  et~al\mbox{.}}{2020}]%
        {song2020learning}
\bibfield{author}{\bibinfo{person}{Hwanjun Song}, \bibinfo{person}{Minseok
  Kim}, \bibinfo{person}{Dongmin Park}, \bibinfo{person}{Yooju Shin}, {and}
  \bibinfo{person}{Jae-Gil Lee}.} \bibinfo{year}{2020}\natexlab{}.
\newblock \showarticletitle{Learning from noisy labels with deep neural
  networks: A survey}.
\newblock \bibinfo{journal}{\emph{arXiv preprint arXiv:2007.08199}}
  (\bibinfo{year}{2020}).
\newblock


\bibitem[\protect\citeauthoryear{Wang, Yang, Ma, Xu, Zhong, Deng, and Gao}{Wang
  et~al\mbox{.}}{2020}]%
        {wang2020weak}
\bibfield{author}{\bibinfo{person}{Yaqing Wang}, \bibinfo{person}{Weifeng
  Yang}, \bibinfo{person}{Fenglong Ma}, \bibinfo{person}{Jin Xu},
  \bibinfo{person}{Bin Zhong}, \bibinfo{person}{Qiang Deng}, {and}
  \bibinfo{person}{Jing Gao}.} \bibinfo{year}{2020}\natexlab{}.
\newblock \showarticletitle{Weak supervision for fake news detection via
  reinforcement learning}. In \bibinfo{booktitle}{\emph{Proceedings of the AAAI
  Conference on Artificial Intelligence}}, Vol.~\bibinfo{volume}{34}.
  \bibinfo{pages}{516--523}.
\newblock


\bibitem[\protect\citeauthoryear{Wardle}{Wardle}{2017}]%
        {wardle2017fake}
\bibfield{author}{\bibinfo{person}{Claire Wardle}.}
  \bibinfo{year}{2017}\natexlab{}.
\newblock \bibinfo{booktitle}{\emph{Fake news. It's complicated}}.
\newblock
\urldef\tempurl%
\url{https://firstdraftnews.org/fake-news-complicated/}
\showURL{%
Retrieved 2019 from \tempurl}


\bibitem[\protect\citeauthoryear{Xie, Dai, Hovy, Luong, and Le}{Xie
  et~al\mbox{.}}{2020}]%
        {xie2020unsupervised}
\bibfield{author}{\bibinfo{person}{Qizhe Xie}, \bibinfo{person}{Zihang Dai},
  \bibinfo{person}{Eduard Hovy}, \bibinfo{person}{Thang Luong}, {and}
  \bibinfo{person}{Quoc Le}.} \bibinfo{year}{2020}\natexlab{}.
\newblock \showarticletitle{Unsupervised Data Augmentation for Consistency
  Training}.
\newblock \bibinfo{journal}{\emph{Advances in Neural Information Processing
  Systems}}  \bibinfo{volume}{33} (\bibinfo{year}{2020}).
\newblock


\bibitem[\protect\citeauthoryear{Yang and Leskovec}{Yang and Leskovec}{2010}]%
        {yang2010modeling}
\bibfield{author}{\bibinfo{person}{Jaewon Yang} {and} \bibinfo{person}{Jure
  Leskovec}.} \bibinfo{year}{2010}\natexlab{}.
\newblock \showarticletitle{Modeling information diffusion in implicit
  networks}. In \bibinfo{booktitle}{\emph{2010 IEEE International Conference on
  Data Mining}}. IEEE, \bibinfo{pages}{599--608}.
\newblock


\bibitem[\protect\citeauthoryear{Zhou, Mulay, Ferrara, and Zafarani}{Zhou
  et~al\mbox{.}}{2020}]%
        {zhou2020recovery}
\bibfield{author}{\bibinfo{person}{Xinyi Zhou}, \bibinfo{person}{Apurva Mulay},
  \bibinfo{person}{Emilio Ferrara}, {and} \bibinfo{person}{Reza Zafarani}.}
  \bibinfo{year}{2020}\natexlab{}.
\newblock \showarticletitle{Recovery: A multimodal repository for covid-19 news
  credibility research}. In \bibinfo{booktitle}{\emph{Proceedings of the 29th
  ACM International Conference on Information \& Knowledge Management}}.
  \bibinfo{pages}{3205--3212}.
\newblock


\bibitem[\protect\citeauthoryear{Zimdars}{Zimdars}{2016}]%
        {zimdars2016false}
\bibfield{author}{\bibinfo{person}{Melissa Zimdars}.}
  \bibinfo{year}{2016}\natexlab{}.
\newblock \bibinfo{booktitle}{\emph{False, Misleading, Clickbait-Y, and
  Satirical `News' Sources}}.
\newblock
\urldef\tempurl%
\url{https://21stcenturywire.com/wp-content/uploads/2017/02/2017-DR-ZIMDARS-False-Misleading-Clickbait-y-and-Satirical-%E2%80%9CNews%E2%80%9D-Sources-Google-Docs.pdf}
\showURL{%
Retrieved 2019 from \tempurl}


\end{thebibliography}

\appendix

\section{Annotation Guidelines}
\label{sec:appx_anno}
\begin{figure}[t]
    \centering
    \begin{subfigure}{\columnwidth}
    \includegraphics[width=\columnwidth]{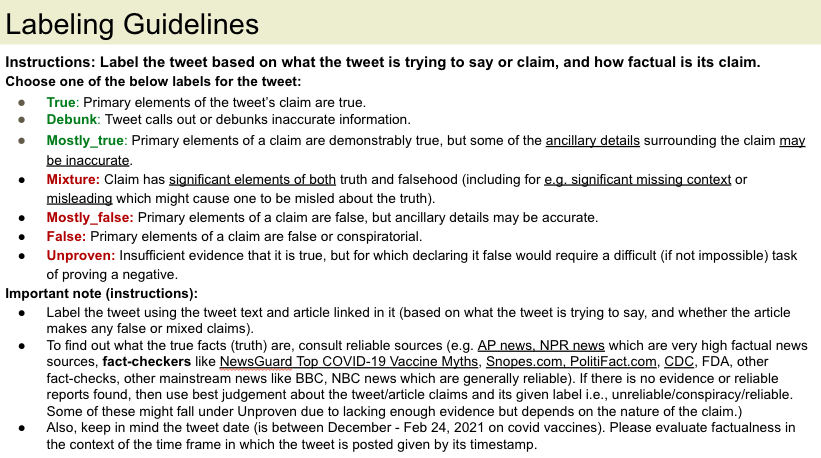}
    \end{subfigure}
    \begin{subfigure}{\columnwidth}
    \includegraphics[width=\columnwidth]{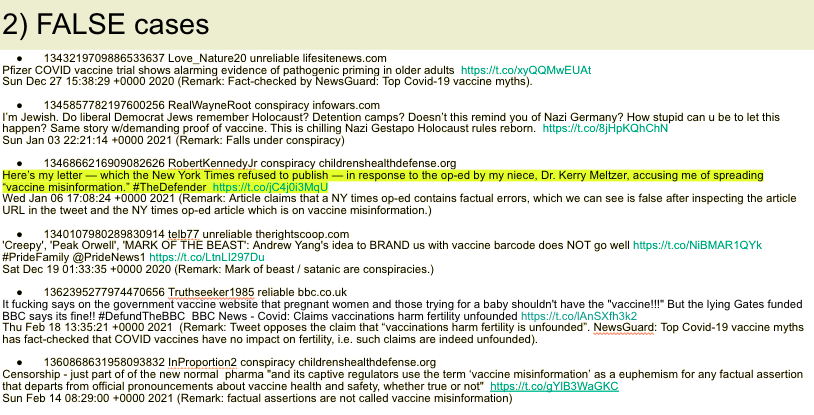}
    \end{subfigure}
    \begin{subfigure}{\columnwidth}
    \includegraphics[width=\columnwidth]{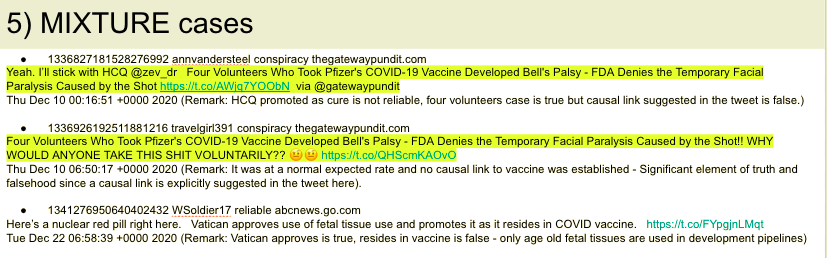}
    \end{subfigure}
    \caption{Annotation guidelines.}
    \label{fig:anno}
\end{figure}

In Fig~\ref{fig:anno}, the instructions and guidelines specified for annotators is included. The annotators are asked to label in the context of when the tweet is posted, with examination of facts from high-factual, low bias news article sources, fact-checking resources, and official information sources. The annotators are provided tweets with the screen name, news source domain, news source label, full tweet text (including the news URL hyperlink), tweet timestamp are provided to aid the annotator. The article URL provides context to the tweet content, and is needed at times to understand the tweet's claim. Typical and trick examples with remarks in each category were provided to review and revisit while annotating, which is a useful guide to provide the distinctions between label types.

\end{document}